\documentclass{elsart}
\usepackage{epsfig}
    \newcommand{\ba}{\begin{eqnarray}}
    \newcommand{\ea}{\end{eqnarray}}
    \newcommand{\be}{\begin{equation}}
    \newcommand{\ee}{\end{equation}}

    \newcommand{\AmS}{{\protect\the\textfont2%
  A\kern-.1667em\lower.5ex\hbox{M}\kern-.125emS}}

\newcommand{\bx}{{\bf x}}

\begin{document}
\runauthor{PKU and ITP and IHEP}
\begin{frontmatter}

\title{Calculating the $I=2$ Pion Scattering Length%
    \ Using Tadpole Improved Clover Wilson Action%
      \ on Coarse Anisotropic Lattices}

\author[PKU]{Chuan Liu}
\author[PKU]{Junhua Zhang}
\author[IHEP]{Ying Chen}
\author[ITP]{and J.~P.~Ma}
\address[PKU]{Department of Physics\\
          Peking University\\
                  Beijing, 100871, P.~R.~China}
\address[ITP]{Institute of Theoretical Physics\\
                Academia Sinica\\
                Beijing, 100080, P.~R.~China}
\address[IHEP]{Institute of High Energy Physics\\
                Academia Sinica\\
                P.~O.~Box 918\\
                Beijing, 100039, P.~R.~China}

\begin{abstract}
In an exploratory study, using the tadpole improved clover Wilson
quark action on small, coarse and anisotropic lattices, the
$\pi\pi$ scattering length in the $I=2$ channel is calculated
within quenched approximation. A new method is proposed which
enables us to make chiral extrapolation of our lattice results
without calculating the decay constant on the lattice. Finite
volume and finite lattice spacing errors are analyzed and the
results are extrapolated towards the infinite volume and
continuum limit. Comparisons of our lattice results with the new
experiment and the results from Chiral Perturbation Theory are
made. Good agreements are found.
\end{abstract}
\end{frontmatter}


\section{Introduction}

 It has become clear that anisotropic, coarse lattices and improved
 lattice actions are ideal candidates for lattice
 QCD calculations on small computers %
 \cite{lepage93:tadpole,lepage95:pc,colin97,colin99}. They are
 particularly advantageous for heavy objects like the glueballs,
 one meson states with nonzero spatial momenta and
 multi-meson states with or without spatial momenta.
 The gauge action employed is the tadpole improved
 gluonic action on anisotropic lattices:
 \ba
 \label{eq:gauge_action}
 S=&-&\beta\sum_{i>j} \left[
 {5\over 9}{TrP_{ij} \over \xi u^4_s}
 -{1\over 36}{TrR_{ij} \over \xi u^6_s}
 -{1\over 36}{TrR_{ji} \over \xi u^6_s} \right] \nonumber \\
 &-&\beta\sum_{i} \left[ {4\over 9}{\xi TrP_{0i} \over  u^2_s}
 -{1\over 36}{\xi TrR_{i0} \over u^4_s} \right] \;\;,
 \ea
 where $P_{0i}$ and $P_{ij}$ represents the usual temporal and
 spatial plaquette variable, respectively.
 $R_{ij}$ and $R_{i0}$ designates the $2\times 1$
 spatial and temporal Wilson loops, where, in order to eliminate
 the spurious states, we have restricted the coupling of
 fields in the temporal direction
 to be within one lattice spacing. The parameter $u_s$,
 which is taken to be the fourth root of the average
 spatial plaquette value,
 is the tadpole improvement parameter determined
 self-consistently from the simulation. With this tadpole
 improvement factor, the renormalization of the anisotropy
 (or aspect ratio) $\xi=a_s/a_t$ will be small. Therefore, we will
 not differentiate the bare aspect ratio and the renormalized one
 in this work.
 Using this action, glueball and hadron spectra have been studied
 within quenched approximation \cite{colin97,colin99,%
 chuan01:gluea,chuan01:glueb,chuan01:canton1,chuan01:canton2,chuan01:india}.
 In this paper, we would like to present our results on
 the pion-pion scattering lengths within quenched approximation.
 Main results of this paper has already been reported
 in \cite{chuan01:i2_letter}. Some details of the
 calculation are presented in this paper.

 Lattice calculations of pion scattering lengths
  have been performed by various authors
 using symmetric lattices without the
 improvement~\cite{gupta93:scat,fukugita95:scat,jlqcd99:scat}.
 It turned out that, using the symmetric lattices and Wilson
 action, large lattices have to be simulated which require
 substantial amount of computing resources.
 It becomes even more challenging if the
 chiral, infinite volume and continuum limits are
 to be studied \cite{jlqcd99:scat}.
 In this exploratory study, we would like to show that, such a calculation
 is feasible using relatively small lattices, typically of
 the size $8^340$, with the
 tadpole improved anisotropic lattice actions.
 Attempts are also made to study the chiral, infinite volume and
 continuum limits extrapolation in a more systematic fashion.
 The final extrapolated result on the pion-pion scattering
 length in the $I=2$ channel is compared with the new
 experimental result from E865 collaboration \cite{experiment_new1:scat}
 and Chiral Perturbation Theory %
 \cite{weinberg:scat,leutwyler83:chiral,bijnens:scat,leutwyler01:scat}
 with encouraging results.
 Of course, more accurate lattice calculations of the pion scattering
 length will inevitably require lattices larger than the ones
 studied in this paper.

 The fermion action used in this calculation is the
 tadpole improved clover Wilson action on anisotropic
 lattices \cite{klassen99:aniso_wilson,chuan01:tune}:
 \ba
 \label{eq:fmatrix}
 {\mathcal M}_{xy} &=&\delta_{xy}\sigma + {\mathcal A}_{xy}
 \nonumber \\
 {\mathcal A}_{xy} &=&\delta_{xy}\left[{1\over 2\kappa_{\max}}
 +\rho_t \sum^3_{i=1} \sigma_{0i} {\mathcal F}_{0i}
 +\rho_s (\sigma_{12}{\mathcal F}_{12} +\sigma_{23}{\mathcal F}_{23}
 +\sigma_{31}{\mathcal F}_{31})\right]
 \nonumber \\
 &-&\sum_{\mu} \eta_{\mu} \left[
 (1-\gamma_\mu) U_\mu(x) \delta_{x+\mu,y}
 +(1+\gamma_\mu) U^\dagger_\mu(x-\mu) \delta_{x-\mu,y}\right] \;\;,
 \ea
 where various coefficients in the fermion matrix
 ${\mathcal M}$ are given by:
 \ba
 \eta_i &=&\nu/(2u_s) \;\;, \;\;
 \eta_0=\xi/2 \;\;, \;\;\sigma=1/(2\kappa)-1/(1\kappa_{max})\;\;,
 \nonumber \\
 \rho_t &=& c_{SW}(1+\xi)/(4u^2_s) \;\;, \;\;
 \rho_s = c_{SW}/(2u^4_s) \;\;.
 \ea
 Among the parameters which appear in the fermion matrix, $c_{SW}$
 is the coefficient of the clover term and $\nu$ is the so-called
 bare velocity of light, which has to be tuned non-perturbatively
 using the single pion dispersion relations \cite{chuan01:tune}.
 Tuning the clover coefficients non-perturbatively is a difficult
 procedure. In this work, only the tadpole improved tree-level
 value is used instead.

 In the fermion matrix~(\ref{eq:fmatrix}), the bare quark mass
 dependence is singled out into the parameter $\sigma$ and the
 matrix ${\mathcal A}$ remains unchanged if the bare quark mass is
 varied.  Therefore, the
 shifted structure of the matrix ${\mathcal M}$ can be utilized
  to solve for quark propagators at various values
  of valance quark mass
 $m_0$ (or equivalently $\kappa$)
 at the cost of solving only the lightest valance
 quark mass value at $\kappa=\kappa_{\max}$,
 using the so-called  Multi-mass Minimal
 Residual ($M^3R$ for short) algorithm %
 \cite{frommer95:multimass,glaessner96:multimass,beat96:multimass}.
 This is particularly advantageous in a quenched calculation
 since one needs the results at various quark mass values
 to perform the chiral extrapolation.

 This paper is organized in the following manner.
 In Section~\ref{sec:theory}, the
 method to calculate $\pi\pi$ scattering length is
 reviewed. In Section~\ref{sec:simulation},
 some simulation details are described.
 In Section~\ref{sec:extrap}, our results
 of the scattering lengths obtained on lattices
 of various sizes and lattice spacings are extrapolated
 towards the chiral limit.
 A new quantity is proposed which is easier to measure
 on the lattice and has a simpler chiral behavior than
 the scattering length itself. Finite size effects
 are studied and two schemes of extrapolating to the
 infinite volume limit are studied. Finally, our lattice
 results are extrapolated towards the continuum limit.
 Comparisons with the experimental value and
 values from chiral perturbation
 theory at various orders are also discussed.
 In Section~\ref{sec:conclude}, we conclude
 with some general remarks.

 \section{Formulation to extract the scattering length}
 \label{sec:theory}

 In order to calculate hadron scattering lengths on the lattice,
 or the scattering phase shifts in general, one uses L\"uscher's
 formula which relates the exact energy level of two hadron states
 in a finite box to the scattering phase shift in the continuum.
 In the case of two pions which scatter at
 zero relative three momentum, this formula then relates the
 {\em exact} two pion energy $E^{(I)}_{\pi\pi}$ in a finite
 box of size $L$ and isospin $I$
 channel to the corresponding scattering length
 $a^{(I)}_0$ in the continuum.
 This formula reads \cite{luscher86:finiteb}:
 \be
 \label{eq:luescher}
 E^{(I)}_{\pi\pi}-2m_\pi=-\frac{4\pi a^{(I)}_0}{m_\pi L^3}
 \left[1+c_1\frac{a^{(I)}_0}{L}+c_2(\frac{a^{(I)}_0}{L})^2
 \right]+O(L^{-6}) \;\;,
 \ee
 where $c_1=-2.837297$, $c_2=6.375183$ are numerical constants.
 In this paper, this formula will be utilized to calculate
 the pion-pion scattering length $a^{(2)}_0$ in the $I=2$ channel,
 which then requires the determination of the corresponding
 energy shift $\delta E^{(2)}_{\pi\pi}\equiv E^{(2)}_{\pi\pi}-2m_\pi$
 in that channel.

 One issue that one has to keep in mind is
 that, in a quenched calculation, L\"uscher's
 formula~(\ref{eq:luescher}) has to be modified
 dramatically as discussed in \cite{bernard96:quenched_scat}.
 Due to the anomalous contributions in quenched chiral
 perturbation theory from $\eta'$, the energy
 shifts are contaminated by terms that might be of order
 $L^0=1$ and $L^{-2}$ in the $I=0$ channel. If these
 contributions were substantial, extraction of the
 corresponding scattering
 length from the energy shift might become problematic. In the
 $I=2$ channel, the situation is better. The quenched chiral
 corrections modify the energy shift by terms that are at least
 of order $L^{-3}$ with usually small coefficients. Therefore,
 doing quenched calculation in the $I=2$ channel is safer.

 To measure the pion mass $m_\pi$ and
 to extract the energy shift $\delta E^{(2)}_{\pi\pi}$
 of two pions with zero relative momentum, we construct the correlation
 functions from the corresponding operators in the $I=2$ channel.
 We have used the operators proposed in
 Ref.~\cite{fukugita95:scat}.
 First, the local operators which create a single pion with
 appropriate isospin values are:
 \ba
 \pi^+(\bx,t)&=&-\bar{d}(\bx,t)\gamma_5u(\bx,t) \;\;,\;\;
 \pi^-(\bx,t)=\bar{u}(\bx,t)\gamma_5d(\bx,t) \;\;,\;\;
 \nonumber \\
 \pi^0(\bx,t)&=&{1 \over \sqrt{2}}
 [\bar{u}(\bx,t)\gamma_5u(\bx,t)-\bar{d}(\bx,t)\gamma_5d(\bx,t)]\;\;,
 \ea
 where $u(\bx,t)$, $d(\bx,t)$, $\bar{u}(\bx,t)$ and
 $\bar{d}(\bx,t)$ are basic local quark fields corresponding to $u$ and
 $d$ quarks respectively.
 The operators which correspond to zero momentum single pions are:
 \be
 \pi^a_0(t)={1 \over L^{3/2}} \sum_{\bx} \pi^a(\bx,t) \;\;,
 \ee
 where the flavor of pions $a=+,-,0$ and $L^3$ is the three volume
 of the lattice. Zero momentum one pion correlation function
 can then be formed as:
 \be
 C_\pi(t)=<\pi^{\bar{a}}_0(t)\pi^a_0(0)> \;\;.
 \ee
 From the large
 $t$ behavior of this correlation function, the pion mass $m_\pi$
 is obtained. Similarly, we use the two pion operators in the
 $I=2$ channel defined by:
 \be
 O^{I=2}_{\pi\pi}(t) = \pi^+_0(t)\pi^+_0(t+1)\;\;,
 \ee
 to construct the two pion correlation functions:
 \be
 C^{I=2}_{\pi\pi}(t)=<O^{I=2}_{\pi\pi}(t)O^{I=2}_{\pi\pi}(0)> \;\;.
 \ee
 From the large $t$ behavior of $C^{I=2}_{\pi\pi}(t)$, one
 could infer the exact two pion state energy in the finite
 box. Numerically, it is more advantageous to construct
 the ratio of the correlation functions defined above:
 \be
 {\mathcal R}^{I=2}(t) = C^{I=2}_{\pi\pi}(t) / (C_\pi(t)C_\pi(t)) \;\;.
 \ee
 This ratio thus exhibits the following asymptotic behavior for
 large $t$:
 \be
 {\mathcal R}(t) \stackrel{t >>1}{\sim} e^{-\delta
 E^{(2)}_{\pi\pi}t} \;\;,
 \ee
 with $\delta E^{(2)}_{\pi\pi}=E^{(2)}_{\pi\pi}-2m_\pi$
 is the energy shift in this channel which directly enters L\"uscher's
 formula~(\ref{eq:luescher}). It is
 argued \cite{sharpe92:scat,fukugita95:scat,bernard96:quenched_scat}
 that this formula can only be utilized for
 small $\delta E^{(2)}_{\pi\pi} t$ values
 in a quenched calculation for large enough $L$.
 In this case, one could equally use the linear fitting function:
 \be
 \label{eq:linear_fit}
 {\mathcal R}(t) \stackrel{T >> t >>1}{\sim} 1-\delta E^{(2)}_{\pi\pi}t
 \;\;,
 \ee
 to determine the energy shift $\delta E^{(2)}_{\pi\pi}$.

 \begin{figure}[htb]
 \begin{center}
 \includegraphics[width=7.0cm, angle=-90]{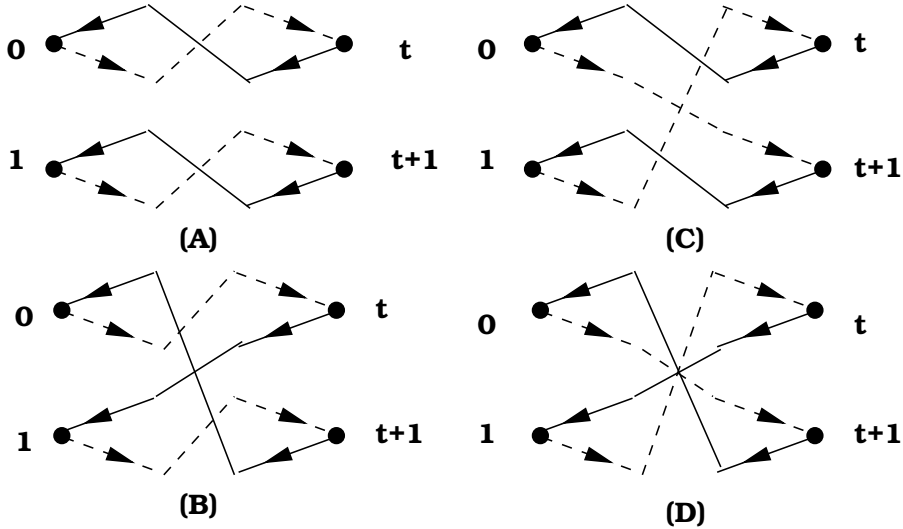}
 \caption{Diagrams representing the contributions to
 the two-pion correlation function
 in the $I=2$ channel. Diagrams (A) and (D) correspond to the
 Direct contribution while (B) and (C) belong to
 the Cross contribution. \label{fig:diagram}}
 \end{center}
 \end{figure}
 Two pion correlation function, or equivalently, the ratio
 ${\mathcal R}(t)$ constructed above can be
 transformed into products of quark propagators using
 Wick's theorem. In order to avoid the complicated
 Fierz re-arrangement terms, we used the creation operators
 at time slices that are differ by one lattice
 spacing as suggested in \cite{fukugita95:scat}. It can be shown that
 the $I=2$ two pion correlation function is given by
 two contributions which are termed Direct and
 Cross contributions \cite{gupta93:scat,fukugita95:scat}.
 These contributions are represented by diagrams
 as shown in Fig.~\ref{fig:diagram}, where the solid
 and dashed lines represent the $u$ and $d$ quark propagators
 respectively.\footnote{In this work, the $u$ and $d$ quark
 are assumed to be degenerate in mass.}%
 The two pion correlation function in the
 $I=0$ channel is, however, more complicated which
 involves vacuum diagrams that require to compute
 the quark propagators for wall sources placed at
 {\em every} time-slice, a procedure which is more
 time-consuming than the $I=2$ channel.
 In the $I=2$ channel, it turns out that
 the quark propagators have to be solved only twice with two
 fixed wall sources placed at $t=0$ and $t=1$.

 \section{Simulation details}
 \label{sec:simulation}

 Simulations are performed on several PC's and two
 workstations.
 Configurations are generated using the pure
 gauge action~(\ref{eq:gauge_action})
 for $4^340$, $6^340$ and $8^340$ lattices with the gauge
 coupling $\beta=1.7$, $2.2$, $2.4$ and $2.6$. The spatial lattice
 spacing $a_s$ is roughly between $0.18$fm and $0.39$fm while
 the physical size of the lattice ranges from
 $0.7$fm to $3.2$fm. For each set of parameters, several hundred
 decorrelated gauge field configurations are used to measure
 the fermionic quantities. Statistical errors are all analyzed
 using the usual jack-knife method. Basic information of these
 configurations is summarized in Table~\ref{tab:parameters}.
 \begin{table}[htb]
 \caption{Simulation parameters for lattices studied in
 this work. Input aspect ratio parameter $\xi$ is fixed to be $5$
 for all lattices being studied.
 The approximate spatial lattice
 spacing $a_s$ in physical units as obtained from
 \cite{colin99,chuan01:india} is also indicated.
 Also listed are the maximum value of the hopping parameter
 $\kappa_{\max}$.
 \label{tab:parameters}}
 \begin{center}
 \begin{tabular}{ccccccc}
 \hline
 $\beta$ & Lattice & $u^4_s$ & $a_s$(fm) & No. confs %
 & $\kappa_{\max}$ & $\nu$ \\
 \hline
 $1.7$ & $4^3\times 40$ & $0.295$ & $0.39$ & $700$ & $0.0570$ & $0.90$\\
 $1.7$ & $6^3\times 40$ & $0.295$ & $0.39$ & $288$ & $0.0585$ & $0.90$\\
 $1.7$ & $8^3\times 40$ & $0.295$ & $0.39$ & $240$ & $0.0585$ & $0.90$\\
 \hline
 $2.2$ & $4^3\times 40$ & $0.378$ & $0.27$ & $700$ & $0.0590$ & $0.95$\\
 $2.2$ & $6^3\times 40$ & $0.378$ & $0.27$ & $256$ & $0.0600$ & $0.95$\\
 $2.2$ & $8^3\times 40$ & $0.378$ & $0.27$ & $224$ & $0.0600$ & $0.95$\\
 \hline
 $2.4$ & $4^3\times 40$ & $0.409$ & $0.22$ & $700$ & $0.0620$ & $0.87$\\
 $2.4$ & $6^3\times 40$ & $0.409$ & $0.22$ & $256$ & $0.0605$ & $0.92$\\
 $2.4$ & $8^3\times 40$ & $0.409$ & $0.22$ & $224$ & $0.0605$ & $0.93$\\
 \hline
 $2.6$ & $4^3\times 40$ & $0.438$ & $0.19$ & $600$ & $0.0630$ & $0.80$\\
 $2.6$ & $6^3\times 40$ & $0.438$ & $0.19$ & $256$ & $0.0620$ & $0.88$\\
 $2.6$ & $8^3\times 40$ & $0.438$ & $0.19$ & $240$ & $0.0605$ & $0.93$\\
 \hline
 \end{tabular}
 \end{center}
 \end{table}

 Quark propagators are measured using the Multi-mass Minimal
 Residue algorithm for $5$ different values of bare quark
 mass. Periodic boundary condition is applied to all three
 spatial directions while in the temporal direction, Dirichlet
 boundary condition is utilized. In this calculation, it is
 advantageous to use the wall sources which greatly enhance
 the signal \cite{gupta93:scat,fukugita95:scat,jlqcd99:scat}.
 Values of the maximum hopping parameter $\kappa_{\max}$,
 which corresponds to the lowest valance quark mass, are also
 listed in Table~\ref{tab:parameters}. Typically, a few hundred
 Minimal Residual iterations are needed to obtain the solution
 vector for a given source vector. On small lattices, in
 particular those with a low value of $\beta$, the hopping
 parameter has to be kept relatively far away from the
 critical kappa value in order to avoid the appearance of
 exceptional configurations.

 \begin{figure}[thb]
 \begin{center}
 \includegraphics[height=12.0cm,angle=0]{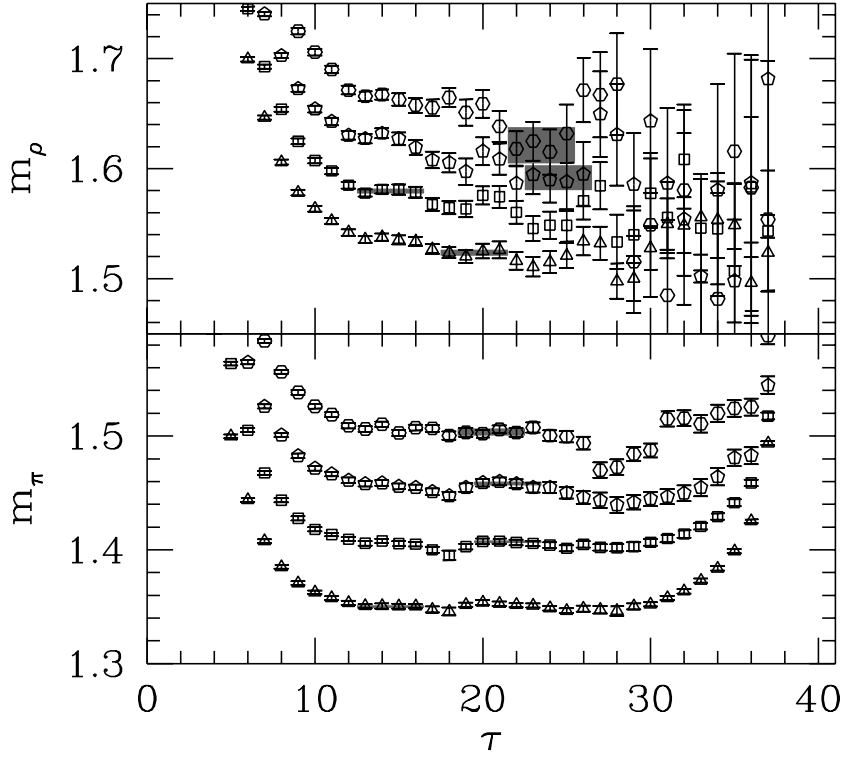}
 \end{center}
 \caption{The single pseudo-scalar (lower) and
 vector (upper) meson effective mass
 plot obtained from the meson correlation functions
 on $8^340$ lattices with $\beta=1.7$. Only the results
 for one value of $\kappa$ is shown. The bare
 velocity of light is taken to be $\nu=0.90$.
 The open triangles, open squares, open pentagons and
 open hexagons are data for the zero and the lowest
 three non-zero spatial lattice
 momenta ($(000)$, $(100)$, $(110)$ and $(111)$),
 respectively. Grey shaded horizontal bars represent
 the fitted values, the errors and the fitting ranges of
 the corresponding energy levels,
 as explained in the text.}
 \label{fig:single_meson}
 \end{figure}
 The single pseudo-scalar and vector meson,
 which will be referred to as pion and rho,
 correlation functions at zero spatial momentum and three lowest lattice
 momenta, namely $(100)$, $(110$ and $(111)$ are constructed from
 the corresponding quark propagators.
 Conventional effective mass functions are computed from which
 the single meson energy levels with zero and non-zero
 momentum are extracted from the effective mass plateaus.
 The fitting range of the plateau is determined automatically
 by requiring the minimum of the $\chi^2$ per degree of freedom of the
 fit. Using the anisotropic lattices, we are
 able to obtain decent effective mass plateaus from the single pion
 and rho correlation functions. In Fig.~\ref{fig:single_meson},
 the effective mass plateaus for the pion and rho are shown
 for one of our parameter set, namely $8^340$ lattices
 with $\beta=1.7$. Points with error bars are
 the effective mass values from the pion (lower half) and
 rho (upper half) correlation functions. For simplicity,
 only data for one particular value of $\kappa$ is shown.
 The grey shaded horizontal regions
 represent the final fitted value of the energy.
 The starting and ending positions of these
 shaded regions correspond to the fitting ranges.
 The height and thickness of each of
 these shaded region designates the fitted
 value and the corresponding error respectively.
 It is seen that single meson energy values
 are obtained with good accuracy, especially for the mass. The
 fitting quality for other parameter sets is similar. The accuracy
 of the mass values are so good that we could
 neglect the errors of these mass values
 when we perform the error analysis of the scattering length
 where the errors from the energy shifts
 $\delta E^{(2)}_{\pi\pi}$ are dominating.

 \begin{figure}[thb]
 \begin{center}
 \includegraphics[height=12.0cm,angle=0]{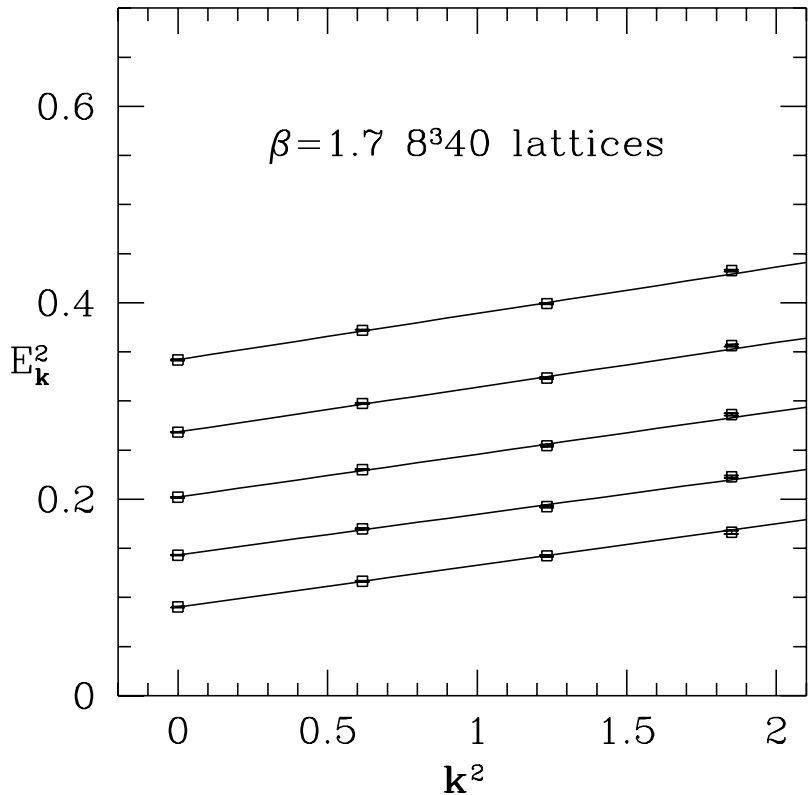}
 \end{center}
 \caption{The dispersion relation for the pseudo-scalar meson
  obtained on $8^340$ lattices with $\beta=1.7$. The bare
 velocity of light is taken to be $\nu=0.90$. It is seen that
 the measured dispersion relations agree with the free dispersion
 relation at small lattice momenta with a slope of $1.06(3)$.}
 \label{fig:dispersion}
 \end{figure}
 Having obtained the single meson energy levels for
 zero and non-zero spatial momentum, the dispersion
 relation of the meson is also at our disposal.
 The parameter $\nu$, also
 known as the bare velocity of light, that enters the fermion
 matrix~(\ref{eq:fmatrix}) is determined non-perturbatively
 using the single pion dispersion relations as described in
 Ref.~\cite{chuan01:tune}. In Fig.~\ref{fig:dispersion},
 we show the single pion dispersion relation obtained at the
 optimal value of $\nu=0.90$ for $\beta=1.7$, $8^340$ lattices.
 Points with error bars are the single pion energy levels
 obtained by fitting the pion effective mass plateaus.
 All data which correspond to $5$ different values of $\kappa$
 are plotted. The straight lines are the linear fits to
 these data. The optimal value of $\nu$ is such that the
 slope of these lines are consistent with unity.
 The optimal value of $\nu$ which corresponds to each
 parameter set of the simulation is also tabulated in
 Table~\ref{tab:parameters}

 \begin{figure}[thb]
 \begin{center}
 \includegraphics[height=12.0cm,angle=0]{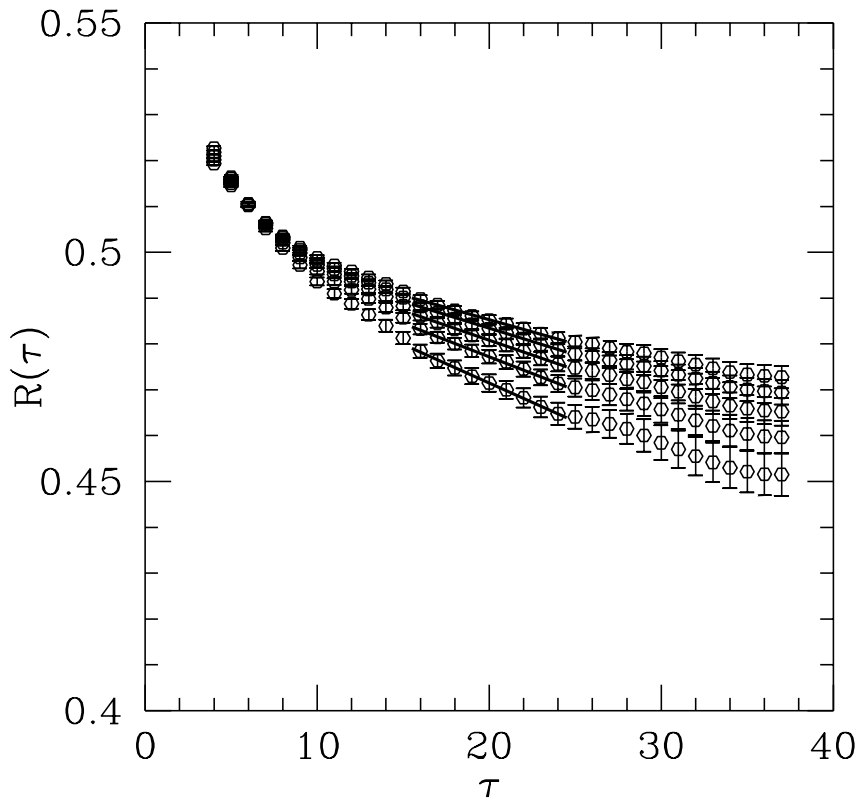}
 \end{center}
 \caption{The ratio $R(t)$ as a function of $t$ obtained from the
 two pion correlation functions at zero lattice momentum
 for $5$ different values of the bare quark mass.
 The lattice size is $8^340$ and the gauge coupling
 $\beta=2.2$. The optimal bare
 velocity of light is taken to be $\nu=0.95$.
 The straight lines are linear fits according to
 Eq.~(\ref{eq:linear_fit}), from which the energy
 shifts $\delta E\equiv E^{(2)}_{\pi\pi}-2m_\pi$ are extracted
 for all $5$ values of $\kappa$.}
 \label{fig:twopion_ratio}
 \end{figure}
 Two pion correlation functions and the ratio ${\mathcal R}(t)$ are
 constructed from products of suitable quark propagators
 according to Wick's theorem. Both the Direct and the Cross
 contributions are included. For the ratio ${\mathcal R}(t)$,
 we obtain good signal for all our data sets.
 In Fig.~\ref{fig:twopion_ratio} we show the ratio
 ${\mathcal R}(t)$ for all five $\kappa$ values
 as a function of the temporal separation $t$,
 together with the corresponding linear fit~(\ref{eq:linear_fit}) for
 $\beta=2.2$ on $8^340$ lattices. The fitting range cannot start
 at a value of $t$ that is too small, particularly in a
 quenched calculation, as discussed in \cite{bernard96:quenched_scat}.
 The starting and ending
 positions of the straight lines indicate the appropriate fitting
 range which minimizes the $\chi^2$ per degree of freedom.
 Here, especially on lattices with larger sizes and better
 signal, we have tried to enlarge the fitting range as much
 as possible. But for small lattices, e.g. $4^340$ lattices
 at $\beta=2.6$, signals only last a few time slices.

 \begin{table}[htb]
 \caption{Extracted values of the single pseudo-scalar,
 vector meson mass and the energy shift $\delta E^{(2)}_{\pi\pi}$
 in the isospin $I=2$ channel obtained from the simulation
 on $4^340$ lattices, all measured
 in $a^{-1}_t$. Also listed are the values of the quantity
 $F={a^{(2)}_0m^2_\rho \over m_\pi}$ (see Sec.~\ref{sec:extrap}
 for the discussion) obtained
 by substituting the value of $\delta E^{(2)}_{\pi\pi}$ into
 Eq.~(\ref{eq:luescher}).
 \label{tab:energies04}}
 \begin{center}
 \begin{tabular}{|c|c|c|c|c|c|c|c|c|c|}
 \hline
 \multicolumn{5}{|c|}{$\beta=1.7$, $4^340$ lattices} &%
 \multicolumn{5}{|c|}{$\beta=2.2$, $4^340$ lattices}  \\
 \hline
 $\kappa$ & $m_\pi$ & $m_\rho$ & %
 $\stackrel{\delta E^{(2)}_{\pi\pi}}{_{(10^{-4})}}$ %
 & $-F$ &%
 $\kappa$ & $m_\pi$ & $m_\rho$ & %
 $\stackrel{\delta E^{(2)}_{\pi\pi}}{_{(10^{-4})}}$ %
 & $-F$ \\
 \hline
 $.057$ & $.4129(6)$ & $.512(2)$  & $99(3)$ & $1.29(3)$&%
 $.059$ & $.275(1)$  & $.334(2)$  & $200(3)$ & $1.04(1)$\\
 $.056$ & $.4827(6)$ & $.570(2)$  & $89(3)$ & $1.42(3)$&%
 $.058$ & $.341(1)$  & $.392(2)$  & $175(3)$ & $1.24(1)$\\
 $.055$ & $.5503(5)$ & $.629(1)$  & $81(3)$ & $1.57(4)$&%
 $.057$ & $.406(1)$  & $.451(1)$  & $157(3)$ & $1.44(1)$\\
 $.054$ & $.6163(5)$ & $.689(1)$  & $75(3)$ & $1.72(4)$&%
 $.056$ & $.4700(9)$ & $.510(1)$  & $143(2)$ & $1.66(2)$\\
 $.053$ & $.6813(5)$ & $.7496(8)$ & $69(2)$ & $1.88(5)$&%
 $.055$ & $.5330(8)$ & $.570(1)$  & $131(2)$ & $1.88(2)$\\
 \hline
 \hline
 \multicolumn{5}{|c|}{$\beta=2.4$, $4^340$ lattices} &%
 \multicolumn{5}{|c|}{$\beta=2.6$, $4^340$ lattices}  \\
 \hline
 $\kappa$ & $m_\pi$ & $m_\rho$ & %
 $\stackrel{\delta E^{(2)}_{\pi\pi}}{_{(10^{-4})}}$ %
 & $-F$ &%
 $\kappa$ & $m_\pi$ & $m_\rho$ & %
 $\stackrel{\delta E^{(2)}_{\pi\pi}}{_{(10^{-4})}}$ %
 & $-F$ \\
 \hline
 $.062$ & $.180(2)$ & $.216(3)$  & $256(19)$ & $0.58(3)$&%
 $.063$ & $.215(1)$  & $.246(2)$  & $143(15)$ & $0.45(4)$\\
 $.061$ & $.245(1)$ & $.276(3)$  & $218(17)$ & $0.78(4)$&%
 $.062$ & $.2701(9)$  & $.286(2)$  & $135(13)$ & $0.56(4)$\\
 $.060$ & $.308(1)$ & $.336(2)$  & $186(18)$ & $0.97(6)$&%
 $.061$ & $.3261(8)$ & $.342(2)$  & $125(12)$ & $0.73(5)$\\
 $.059$ & $.368(1)$ & $.395(2)$  & $140(10)$ & $1.03(5)$&%
 $.060$ & $.3821(8)$ & $.398(2)$  & $116(11)$ & $0.89(6)$\\
 $.058$ & $.428(1)$ & $.453(2)$ & $158(21)$ & $1.45(12)$&%
 $.059$ & $.4380(7)$ & $.454(1)$  & $107(10)$ & $1.06(7)$\\
 \hline
 \end{tabular}
 \end{center}
 \end{table}
 \begin{table}[htb]
 \caption{Same as Fig.~\ref{tab:energies04} but for
 $6^340$ lattices.
 \label{tab:energies06}}
 \begin{center}
 \begin{tabular}{|c|c|c|c|c|c|c|c|c|c|}
 \hline
 \multicolumn{5}{|c|}{$\beta=1.7$, $6^340$ lattices} &%
 \multicolumn{5}{|c|}{$\beta=2.2$, $6^340$ lattices}  \\
 \hline
 $\kappa$ & $m_\pi$ & $m_\rho$ & %
 $\stackrel{\delta E^{(2)}_{\pi\pi}}{_{(10^{-4})}}$ %
 & $-F$ &%
 $\kappa$ & $m_\pi$ & $m_\rho$ & %
 $\stackrel{\delta E^{(2)}_{\pi\pi}}{_{(10^{-4})}}$ %
 & $-F$ \\
 \hline
 $.0585$ & $.3029(8)$ & $.416(2)$  & $43(6)$ & $1.32(14)$&%
 $.060$  & $.202(1)$  & $.294(2)$  & $81(6)$ & $1.19(7)$\\
 $.0575$ & $.3803(7)$ & $.477(1)$  & $37(5)$ & $1.45(15)$&%
 $.059$  & $.2751(7)$ & $.348(2)$  & $66(5)$ & $1.33(8)$\\
 $.0565$ & $.4472(5)$ & $.537(1)$  & $32(4)$ & $1.62(16)$&%
 $.058$  & $.3440(8)$ & $.404(1)$  & $56(5)$ & $1.51(9)$\\
 $.0555$ & $.5159(4)$ & $.597(1)$  & $30(4)$ & $1.82(18)$&%
 $.057$  & $.4097(7)$ & $.461(1)$  & $49(5)$ & $1.72(12)$\\
 $.0545$ & $.5827(4)$ & $.657(1)$  & $28(4)$ & $2.03(20)$&%
 $.056$  & $.4737(6)$ & $.520(1)$  & $43(4)$ & $1.91(13)$\\
 \hline
 \hline
 \multicolumn{5}{|c|}{$\beta=2.4$, $6^340$ lattices} &%
 \multicolumn{5}{|c|}{$\beta=2.6$, $6^340$ lattices}  \\
 \hline
 $\kappa$ & $m_\pi$ & $m_\rho$ & %
 $\stackrel{\delta E^{(2)}_{\pi\pi}}{_{(10^{-4})}}$ %
 & $-F$ &%
 $\kappa$ & $m_\pi$ & $m_\rho$ & %
 $\stackrel{\delta E^{(2)}_{\pi\pi}}{_{(10^{-4})}}$ %
 & $-F$ \\
 \hline
 $.0605$ & $.205(1)$  & $.267(3)$  & $66(21)$ & $0.82(20)$&%
 $.0615$ & $.199(1)$  & $.232(2)$  & $130(10)$ & $1.09(6)$\\
 $.0595$ & $.275(1)$  & $.322(2)$  & $62(8)$  & $1.08(10)$&%
 $.0605$ & $.2626(9)$ & $.291(1)$  & $106(8)$ & $1.38(6)$\\
 $.0585$ & $.3402(9)$ & $.379(1)$  & $48(10)$ & $1.18(19)$&%
 $.0595$ & $.3241(6)$ & $.351(1)$  & $91(7)$  & $1.70(8)$\\
 $.0575$ & $.4030(8)$ & $.436(1)$  & $38(10)$ & $1.25(25)$&%
 $.0585$ & $.3840(5)$ & $.409(2)$  & $79(6)$  & $2.00(10)$\\
 $.0565$ & $.4610(6)$ & $.494(2)$  & $31(10)$ & $1.30(32)$&%
 $.0575$ & $.4439(6)$ & $.467(1)$  & $71(6)$  & $2.30(12)$\\
 \hline
 \end{tabular}
 \end{center}
 \end{table}
 \begin{table}[htb]
 \caption{Same as Fig.~\ref{tab:energies04} but for
 $8^340$ lattices.
 \label{tab:energies08}}
 \begin{center}
 \begin{tabular}{|c|c|c|c|c|c|c|c|c|c|}
 \hline
 \multicolumn{5}{|c|}{$\beta=1.7$, $8^340$ lattices} &%
 \multicolumn{5}{|c|}{$\beta=2.2$, $8^340$ lattices}  \\
 \hline
 $\kappa$ & $m_\pi$ & $m_\rho$ & %
 $\stackrel{\delta E^{(2)}_{\pi\pi}}{_{(10^{-4})}}$ %
 & $-F$ &%
 $\kappa$ & $m_\pi$ & $m_\rho$ & %
 $\stackrel{\delta E^{(2)}_{\pi\pi}}{_{(10^{-4})}}$ %
 & $-F$ \\
 \hline
 $.0585$ & $.3005(7)$ & $.421(2)$  & $21(3)$ & $1.57(21)$&%
 $.060$  & $.2026(7)$ & $.279(2)$  & $45(9)$ & $1.43(23)$\\
 $.0575$ & $.3782(6)$ & $.481(1)$  & $14(2)$ & $1.43(13)$&%
 $.059$  & $.2758(6)$ & $.338(1)$  & $37(8)$ & $1.66(25)$\\
 $.0565$ & $.4495(5)$ & $.542(1)$  & $13(1)$ & $1.62(14)$&%
 $.058$  & $.3432(5)$ & $.397(1)$  & $32(7)$ & $1.97(29)$\\
 $.0555$ & $.5180(4)$ & $.597(1)$  & $12(1)$ & $1.78(17)$&%
 $.057$  & $.4084(4)$ & $.4558(6)$ & $29(6)$ & $2.30(34)$\\
 $.0545$ & $.5846(4)$ & $.657(1)$  & $11(1)$ & $1.98(18)$&%
 $.056$  & $.4721(4)$ & $.5150(6)$ & $26(6)$ & $2.65(39)$\\
 \hline
 \hline
 \multicolumn{5}{|c|}{$\beta=2.4$, $8^340$ lattices} &%
 \multicolumn{5}{|c|}{$\beta=2.6$, $8^340$ lattices}  \\
 \hline
 $\kappa$ & $m_\pi$ & $m_\rho$ & %
 $\stackrel{\delta E^{(2)}_{\pi\pi}}{_{(10^{-4})}}$ %
 & $-F$ &%
 $\kappa$ & $m_\pi$ & $m_\rho$ & %
 $\stackrel{\delta E^{(2)}_{\pi\pi}}{_{(10^{-4})}}$ %
 & $-F$ \\
 \hline
 $.0605$ & $.1966(7)$ & $.256(1)$   & $56(6)$  & $1.43(11)$&%
 $.0605$ & $.1991(5)$ & $.2394(9)$  & $64(10)$ & $1.39(15)$\\
 $.0595$ & $.2662(6)$ & $.3130(8)$  & $42(6)$  & $1.59(15)$&%
 $.0595$ & $.2652(5)$ & $.2988(7)$  & $52(7)$  & $1.73(15)$\\
 $.0585$ & $.3317(5)$ & $.3717(6)$  & $34(5)$  & $1.83(19)$&%
 $.0585$ & $.3285(4)$ & $.3579(6)$  & $42(9)$  & $2.03(27)$\\
 $.0575$ & $.3950(5)$ & $.4303(6)$  & $29(4)$  & $2.09(20)$&%
 $.0575$ & $.3904(3)$ & $.4167(6)$  & $36(8)$  & $2.33(33)$\\
 $.0565$ & $.4569(4)$ & $.4887(5)$  & $25(5)$  & $2.34(31)$&%
 $.0565$ & $.4513(3)$ & $.4753(5)$  & $32(7)$  & $2.65(39)$\\
 \hline
 \end{tabular}
 \end{center}
 \end{table}
 Finally, the extracted values of the single meson energy and the
 energy shifts are all listed in
 Table~\ref{tab:energies04},Table~\ref{tab:energies06}
 and Table~\ref{tab:energies08} for $4^340$, $6^340$ and
 $8^340$ lattices, respectively.

 After obtaining the energy shifts $\delta E^{(2)}_{\pi\pi}$,
 these values are substituted into L\"uscher's formula to solve
 for the scattering length $a^{(2)}_0$ for all five $\kappa$
 values. This is done for lattices of all sizes being
 simulated and for all values of $\beta$.
 The results for the quantity $F\equiv a^{(2)}_0m^2_\rho/m_\pi$ (see the next
 section for the reason of this choice) are
 also listed in Table~\ref{tab:energies04} to
 Table~\ref{tab:energies08}.
 From these results, attempts are made to perform
 an extrapolation towards the chiral, infinite volume
 and zero lattice spacing limit.

 \section{Extrapolation towards chiral, infinite volume and continuum limits}
 \label{sec:extrap}

 Since the valance quark mass values being studied are
 far from the chiral limit, we first try to make the
 chiral extrapolation at finite volume and finite lattice
 spacing. This is facilitated by results for the scattering
 length at $5$ different values of $\kappa$.
 In the chiral limit, the $\pi\pi$ scattering length in
 the $I=2$ channel is given by the current algebra
 result due to S.~Weinberg \cite{weinberg:scat}:
 \be
 \label{eq:weinberg}
 a^{(2)}_0 =-{1 \over 16 \pi} {m_\pi \over f^2_\pi} \;\;,
 \ee
 where $a^{(2)}_0$ is the $\pi\pi$ scattering length $a_0$ in
 the $I=2$ channel and $f_\pi\sim 93$MeV is the pion decay constant.
 Chiral Perturbation Theory to one-loop order gives an
 expression which includes the next-to-leading order
 contributions \cite{leutwyler83:chiral}. Even the
 next-next-to-leading order corrections have been
 calculated within Chiral Perturbation
 Theory \cite{bijnens:scat}. For a summary of the
 results in Chiral Perturbation Theory, see
 Ref.~\cite{leutwyler01:scat} and the references therein.
 Since the one-loop and two-loop {\em numerical} results on the pion-pion
 scattering length in the $I=2$ channel do not differ from
 the current algebra value substantially, we will mainly
 compare with Weinberg's value.
 Complication arises in the quenched approximation. In principle,
 the quenched scattering lengths becomes divergent in the
 chiral limit \cite{bernard96:quenched_scat}. However, these
 divergent terms only become numerically important when the
 pion mass is close to zero. For the parameters used in our
 simulation, these terms seem to be numerically small in the
 $I=2$ channel and we could not observe this divergence
 from our data.

 In most of the previous lattice
 calculations \cite{gupta93:scat,fukugita95:scat,jlqcd99:scat},
 both the mass and the decay constant of the pion
 were calculated on the lattice. Then, the lattice
 results of $m_\pi$ and $f_\pi$ were substituted
 into the current algebra result~(\ref{eq:weinberg})
 to obtain a prediction of the scattering length.
 This is to be compared with the scattering length
 obtained from L\"uscher's formula. In these studies,
 some discrepancies between the lattice results and
 the chiral results was observed \cite{jlqcd99:scat}.
 There were also
 discrepancies between the staggered fermion results
 and the Wilson fermion results. The major disadvantage
 of the this procedure is the following: First, it is
 difficult to make a direct chiral extrapolation towards
 the chiral limit; Second, the lattice results of the
 decay constant are usually much less accurate,
 both statistically and systematically, than the
 mass values. In fact, most of the discrepancies are
 due to inaccuracy of the decay constants, as
 realized already in \cite{jlqcd99:scat}.
 In this paper, we propose to use
 another quantity %
 \footnote{A similar dimensionful, instead of
 dimensionless, quantity $a^{(2)}_0/m_\pi$has been
 proposed in Ref.\cite{jlqcd99:scat}.}
 which has a simpler chiral behavior
 to compare with the current algebra result~(\ref{eq:weinberg}).
 This quantity is simply $F=a^{(2)}_0m^2_\rho/m_\pi$, which in
 the chiral limit reads:
 \be
 \label{eq:chiral}
 F\equiv {a^{(2)}_0m^2_\rho \over m_\pi}
 =-{1 \over 16 \pi} {m^2_\rho \over f^2_\pi}
 \sim  -1.3638\;\;,
 \ee
 where the final numerical value is obtained
 by substituting in the experimental values for
 $m_\rho\sim 770$MeV and $f_\pi\sim 93$MeV.
 Note that both $m_\rho$ and
 $f_\pi$ have {\em finite} values when
 approaching the chiral limit. Therefore, the above
 combination will approach a finite value in the
 chiral limit. On the lattice, we could extract the scattering
 length $a^{(2)}_0$ from L\"uscher's formula. More importantly,
 the mass of the pion and the rho meson can be obtained with
 good accuracy on the lattice. So, the factor $F$
 can be calculated on the lattice with good
 precision {\em without} the
 lattice calculation of $f_\pi$. The error
 of the factor $F$ obtained on the lattice will mainly
 come from the error of the scattering length $a^{(2)}_0$, or
 equivalently, the energy shift
 $\delta E^{(2)}_{\pi\pi}$ and nowhere else.
 Since we have calculated the factor $F$ for $5$ different
 values of valance quark mass, we could make a chiral
 extrapolation and extract the result in the chiral limit.
 Comparisons with Weinberg's result~(\ref{eq:chiral}) and
 the experiment will offer us a cross check among different
 methods.

 \begin{figure}[thb]
 \begin{center}
 \includegraphics[height=12.0cm,angle=0]{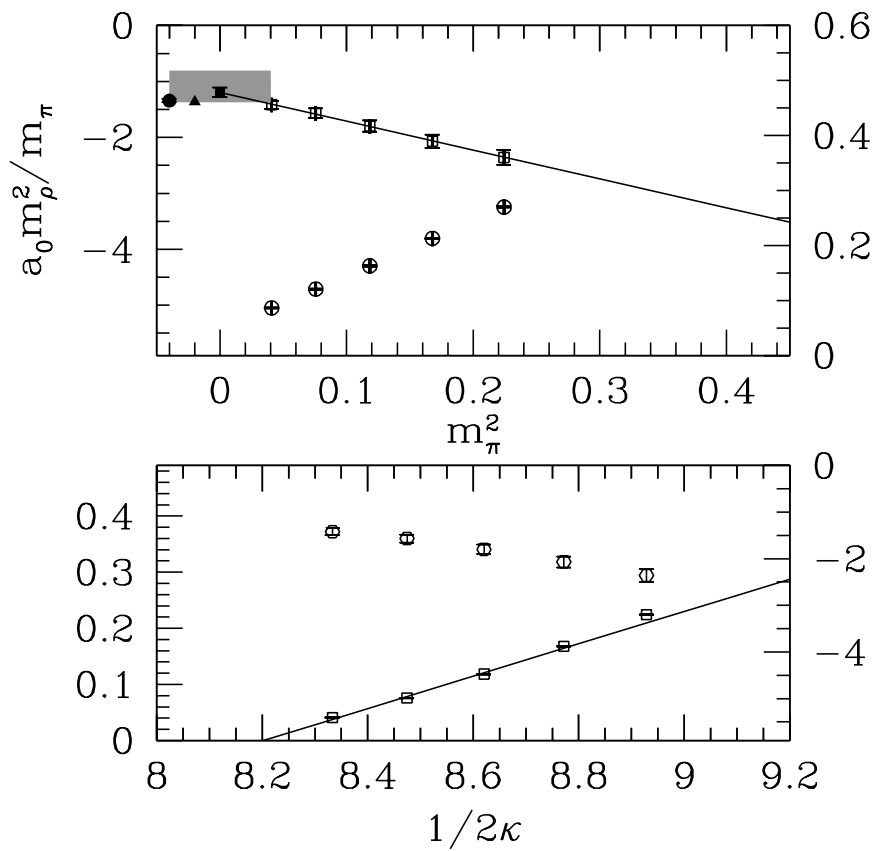}
 \end{center}
 \caption{Chiral extrapolation for the quantity
 $F\equiv a^{(2)}_0m^2_\rho/m_\pi$ for our simulation results
 at $\beta=2.2$ on $6^340$ lattices. In the lower
 half of the plot, the pseudo-scalar
 mass squared are plotted as open squares as function of
 $1/(2\kappa)$. The straight line represents
 the corresponding linear fit~(\ref{eq:mq}) for the data.
 Also shown in the lower half as open hexagons are the results
 for the factor $F$. In the upper half
 of the plot, the same quantity $F$ is plotted (open squares)
 as a function of $m^2_\pi$. Also plotted as open circles are the data
 points for $m^2_\rho$. The straight line represents the linear
 extrapolation towards the chiral limit $m^2_\pi=0$, where
 the extrapolated result is also depicted as a solid square.
 As a comparison, the
 corresponding experimental result \cite{experiment_new1:scat}
 for this quantity is drawn as a shaded band. Weinberg's result
 and the result from chiral perturbation theory are also shown
 as a filled triangle and a filled circle at $m_\pi=0$, respectively.
 \label{fig:chiral_extrapolation}}
 \end{figure}
 In the upper half of Fig.~\ref{fig:chiral_extrapolation},
 we show the chiral extrapolation of the quantity
 $F=a^{(2)}_0m^2_\rho/m_\pi$ as a function of the pseudo-scalar
 mass squared ($m^2_\pi$) for the simulation
 on  $6^340$ lattices at $\beta=2.2$. The  open
 squares represent data points for $F$ at various values
 of $m^2_\pi$. The straight
 line is the linear chiral extrapolation and the extrapolated result
 is also depicted as a solid square at $m_\pi=0$.
 Weinberg's result
 and the result from chiral perturbation theory are also shown
 as a filled triangle and a filled circle at $m_\pi=0$, respectively.
 The vector meson mass values squared are
 also shown as the open circles in the upper half of this plot.
 The pseudo-scalar
 meson mass squared $m^2_\pi$ (open squares)
 are plotted in the lower half of the figure as
 a function of $1/(2\kappa)$, which linearly depends
 on the valance quark mass $m_q$ via:
 \be
 \label{eq:mq}
 m_q=A\left({1\over 2\kappa}-{1\over 2\kappa_c}\right)
 \;\;,
 \ee
 It is seen that meson mass squared depends on
 the valance quark mass linearly, as expected
 from chiral symmetry. It is also possible to dig
 out the critical value of the hopping parameter
 where the pseudo-scalar mass vanishes. The data points
 for the factor $F$ are also shown in the lower half
 of the plot as open circles.
 The fitting quality for the pion, rho
 and the factor $F$ are reasonable.
 The quality for the chiral extrapolation of other simulation points
 are similar. As is seen, the linear fit gives a
 reasonable modeling of the data. The divergent contributions
 from quenched chiral perturbation theory seem to be
 numerically small for the lattices being simulated.

 \begin{figure}[thb]
 \begin{center}
 \includegraphics[height=12.0cm,angle=0]{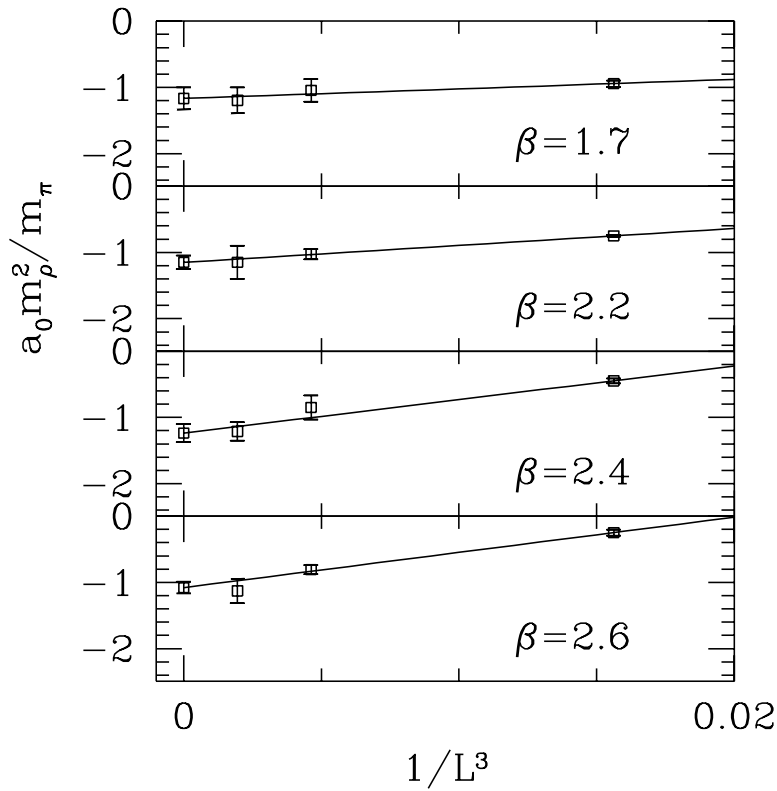}
 \end{center}
 \caption{Infinite volume extrapolation in Scheme I for the quantity
 $F=a^{(2)}_0m^2_\rho/m_\pi$ for our simulation results
 at $\beta=2.6$, $2.4$, $2.2$ and $1.7$. The straight lines represent
 the corresponding linear extrapolation in $1/L^3$.
 The extrapolated result is also shown, together with its error.}
 \label{fig:extrapL}
 \end{figure}
 \begin{figure}[htb]
 \begin{center}
 \includegraphics[height=12.0cm,angle=0]{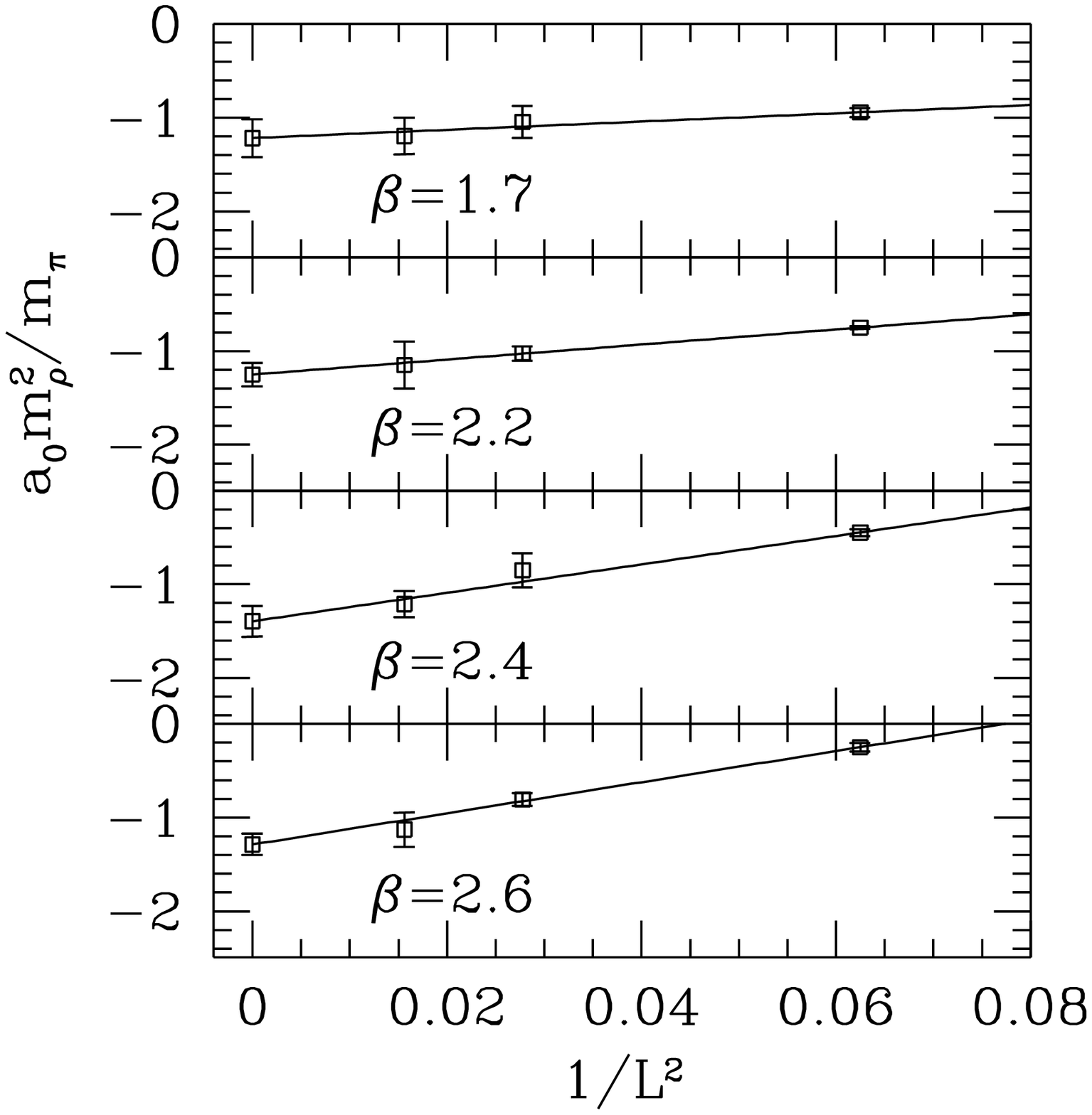}
 \end{center}
 \caption{Same as Fig.~\ref{fig:extrapL} but
 for Scheme II.
 \label{fig:extrapL2}}
 \end{figure}
 After the chiral extrapolation, we now turn to study
 the finite volume effects of the simulation. According
 to formula~(\ref{eq:luescher}), the quantity
 $F$ obtained from finite lattices differ from
 its infinite volume value by corrections of the
 form $1/L^3$. However, it was argued in
 Ref.\cite{sharpe92:scat,fukugita95:scat,bernard96:quenched_scat}
 that in a quenched calculation, the form of L\"uscher's
 formula is invalidated. Finite volume corrections will
 have a different dependence on the volume, e.g. a correction
 of the form $1/L^5$ instead of $1/L^6$, as predicted
 by formula~(\ref{eq:luescher}). This would mean that the
 factor $F$ receives finite volume
 correction of the form $1/L^2$. In our simulation, however,
 we were unable to judge from our data which extrapolation
 is more convincing. We therefore perform our infinite volume
 extrapolation in both ways, calling them scheme I ( extrapolating
 according to $1/L^3$ ) and scheme II ( extrapolating according to
 $1/L^2$ ), respectively. In fact,
 extrapolation in these two different
 schemes yields compatible results within
 statistical errors. The fitting quality of Scheme II
 is somewhat, but not overwhelmingly, better than that of
 Scheme I.
 In Fig.~\ref{fig:extrapL} and Fig.~\ref{fig:extrapL2},
 we have shown the infinite volume extrapolation according to
 Scheme I and Scheme II, respectively, for the simulation points
 at $\beta=2.6$, $2.4$, $2.2$ and $1.7$. The extrapolated
 results are shown with open squares at
 $L=\infty$, together with the corresponding errors. The straight lines
 represent the linear extrapolation in $1/L^3$ or $1/L^2$.
 It is seen that, on physically small lattices, e.g. those with
 $\beta=2.6$ and $\beta=2.4$, the finite volume correction
 is much more significant than larger lattices. For the
 lattices with $\beta=1.7$, the finite volume dependence of
 the result is rather weak, showing that even on
 $4^340$ lattices with this lattice spacing, the finite
 volume correction is small. This is also reflected by
 the slopes of the linear fits. It is seen from
 the figures that the slopes of the fitting straight lines
 increase with $\beta$, showing a larger finite volume
 correction at larger $\beta$ values hence smaller lattices.

 \begin{figure}[thb]
 \begin{center}
 \includegraphics[height=12.0cm,angle=0]{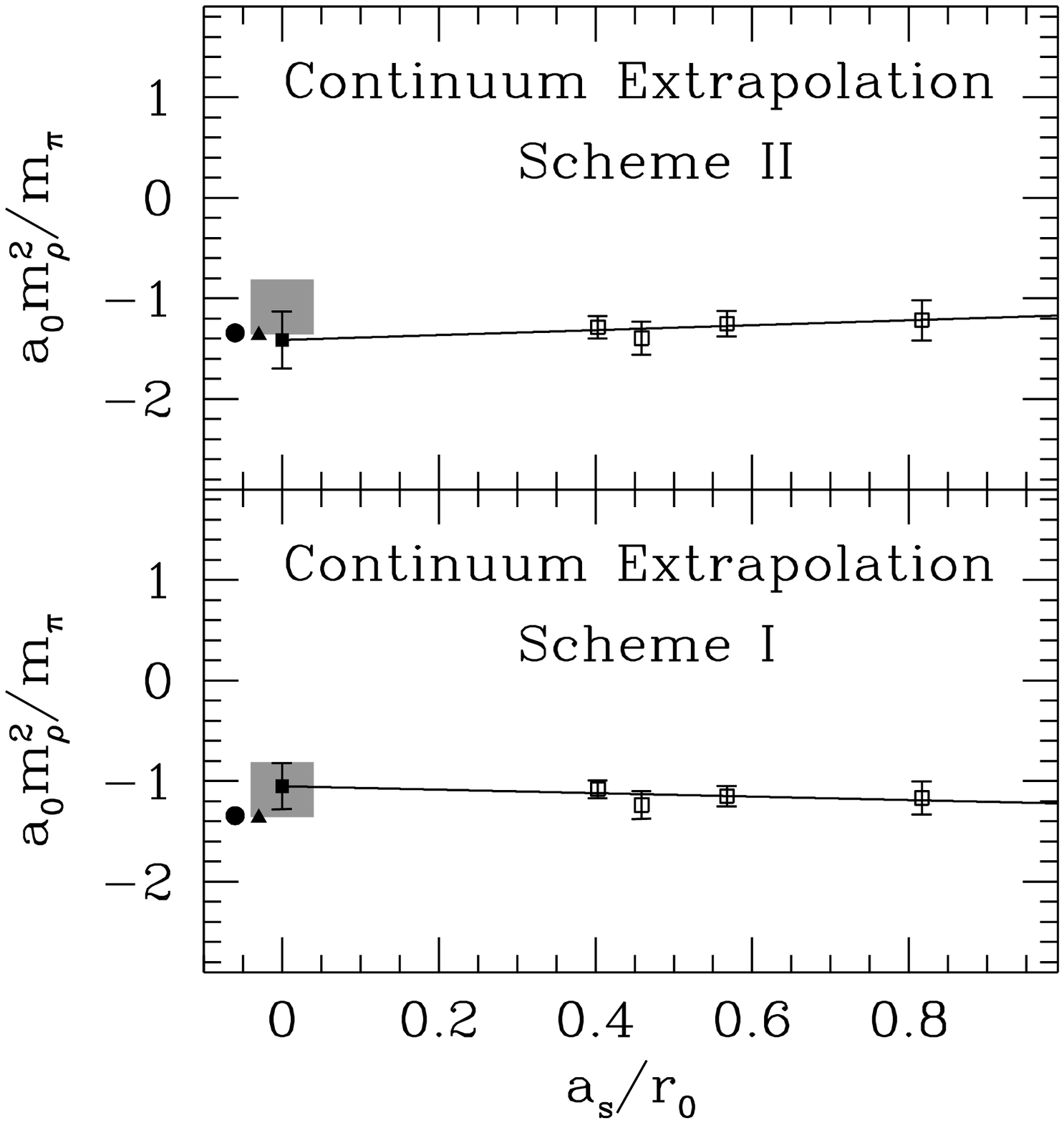}
 \end{center}
 \caption{Continuum extrapolation for the quantity
 $a^{(2)}_0m^2_\rho/m_\pi$ obtained from our simulation results
 at $\beta=2.6$, $2.4$, $2.2$ and $1.7$. The results for both
 Scheme I and II are shown. The straight lines represent
 the linear extrapolation in $a_s/r_0$.
 The extrapolated results are also shown, together with the
 experimental result from Ref.~\cite{experiment_new1:scat} indicated
 by the shaded band. For comparison, Weinberg's
 result~(\ref{eq:chiral}) and the results from Chiral perturbation
 theory are also shown as the points at $a_s=0$ respectively.
 \label{fig:continuum_extrapolation}}
 \end{figure}
 Finally, we can make an extrapolation towards the
 continuum limit by eliminating the finite lattice
 spacing errors. Since we have used the tadpole
 improved clover Wilson action, all physical quantities
 differ from their continuum counterparts by
 terms that are proportional to $a_s$. The physical
 value of $a_s$ for each value of $\beta$ can be
 found from Ref. \cite{colin99,chuan01:india}, which
 is also included in Table~\ref{tab:parameters}.
 This extrapolation is shown in
 Fig.~\ref{fig:continuum_extrapolation} where
 the results from the chiral and infinite volume
 extrapolation discussed above are indicated as
 data points in the plot for all $4$ values of
 $\beta$ that have been simulated. In the lower/upper half
 of the plot, results in Scheme I/II are shown.
 The straight lines show
 the extrapolation towards the $a_s=0$ limit and the
 extrapolated results are also shown as solid squares together with
 the experimental result from Ref.\cite{experiment_new1:scat}
 which is shown as the shaded band. For comparison with
 chiral perturbation theory, Weinberg's
 result~(\ref{eq:chiral}) and the results from Chiral perturbation
 theory are also shown as the filled triangles and
 filled circles at $a_s=0$, respectively.
 It is seen that
 our lattice calculation gives a compatible result
 for the quantity $a^{(2)}_0m^2_\rho/m_\pi$ when compared
 with the experiment. The statistical errors for the
 final result is still somewhat large. This is mainly
 due to lack of results at smaller lattice spacings.
 However, the result is promising since the chiral,
 infinite volume and continxuum extrapolation for
 the pion scattering length have not been studied
 systematically using small lattices before.
 Another encouraging sign
 is that all data points at different lattice
 spacing values are about the same which indicates
 that the $O(a_s)$ lattice
 effects are small, presumably due to the tadpole improvement of
 the action. To summarize,
 we obtain from the linear extrapolation the following
 result for the quantity $F=a^{(2)}_0m^2_\rho/m_\pi$:
 \ba
 {a^{(2)}_0m^2_\rho \over m_\pi}&=&-1.05(23)
 \;\;\mbox{for Scheme I}\;\;,
 \nonumber \\
 {a^{(2)}_0m^2_\rho \over m_\pi}&=&-1.41(28)
 \;\;\mbox{for Scheme II}\;\;.
 \ea
 If we substitute in the mass of the mesons from
 the experiment: $m_\rho\sim 770$MeV and $m_\pi\sim 139$MeV,
 we obtain the quantity $a^{(2)}_0m_\pi$:
 \ba
 a^{(2)}_0m_\pi&=&-0.0342(75) \;\;\mbox{for Scheme I}\;\;,
 \nonumber \\
 a^{(2)}_0m_\pi&=&-0.0459(91) \;\;\mbox{for Scheme II}\;\;.
 \ea
 Weinberg's current algebra prediction~(\ref{eq:weinberg})
 yields a value of $a^{(2)}_0m_\pi=-0.046$.
 This quantity has been calculated in Chiral Perturbation
 Theory to one-loop order with the result:
 $a^{(2)}_0m_\pi=-0.042$ \cite{leutwyler83:chiral} and
 recently to two-loop order \cite{bijnens:scat,leutwyler01:scat}.
 The final result from Chiral Perturbation Theory
 gives: $a^{(2)}_0m_\pi=-0.0444(10)$, where the
 error comes from theoretical uncertainties.
 On the experimental side, a new
 result from E865 collaboration
 claims $a^{(2)}_0m_\pi=-0.036(9)$. It is encouraging to find out
 that our lattice results in {\em both} schemes are compatible with
 the experiment. Our result in Scheme II also agrees with the
 Chiral Perturbation Theory results very well while our result
 in Scheme I is barely within one standard deviation of the chiral
 results.

 \section{Conclusions}
 \label{sec:conclude}

 In this paper, we have calculated pion-pion scattering
 lengths in isospin $I=2$ channel using quenched
 lattice QCD. It is shown that such a calculation
 is feasible using relatively small, coarse and anisotropic lattices
 with limited computer resources like several
 personal computers and workstations.
 The calculation is done using the
 tadpole improved clover Wilson action on anisotropic
 lattices. Simulations are performed on lattices
 with various sizes, ranging from $0.8$fm to about
 $3$fm and with different value of lattice spacing.
 Quark propagators are measured with $5$ different
 valance quark mass values. These enable us to
 explore the finite volume errors and the finite
 lattice spacing errors in a more systematic fashion.
 The infinite volume extrapolation is explored in
 two different schemes which yields compatible final
 results. The lattice result for the scattering length is
 extrapolated towards the chiral
 and continuum limit where a result consistent with
 the experiment and the Chiral Perturbation Theory is found.
 We believe that, using the method described in this exploratory
 study, more reliable and accurate results on pion
 scattering lengths could be obtained
 with simulations on larger lattices and computers.

 Finally, our method for calculating the pion-pion
 scattering length discussed in this paper can easily
 be generalized to calculate the scattering lengths
 of other hadrons, or in other channels, e.g. $I=0$ channel,
  where extra care has to be taken due to enhanced terms
 coming from quenched chiral loops.
 The method can also be applied to calculate the scattering
 phase shift at non-zero spatial lattice momenta, where
 presumably lattices larger than the ones utilized in
 this exploratory study is needed.

 \section*{Acknowledgments}

 This work is supported by the National Natural
 Science Foundation of China (NSFC) under grant
 No. 90103006 and Pandeng fund. C. Liu would like
 to thank Prof.~H.~Q.~Zheng for helpful discussions.
 The authors would like to thank Prof. C.~Bernard for
 bringing our attention to the issue of enhanced
 quenched chiral loop contributions.


\end{document}